\input ./style/arxiv-general.cfg
\documentclass[MSNbibl,nameyear,dvips]{arxstspdf}
\makeatletter
   \@ifpackageloaded{graphicx}{}{\usepackage{graphicx}}
\makeatother
\usepackage{flushend}
\usepackage{stfloats}

\volume{30}
\issue{3}
\pubyear{2015}
\firstpage{413}
\lastpage{422}
\doi{10.1214/15-STS521}
\docsubty{FLA}


\begin{document}
\begin{frontmatter}

\title{A Conversation with Alan Gelfand}
\runtitle{A Conversation with Alan Gelfand}

\begin{aug}
\author[A]{\fnms{Bradley P.} \snm{Carlin}\corref{}\ead[label=e1]{brad@biostat.umn.edu}}
\and
\author[B]{\fnms{Amy H.} \snm{Herring}\ead[label=e2]{aherring@bios.unc.edu}}
\runauthor{B.~P. Carlin and A.~H. Herring}

\affiliation{University of Minnesota and University of North Carolina
at Chapel Hill}

\address[A]{Bradley P. Carlin is Professor and Head of Biostatistics, Division of Biostatistics, School of Public Health,
University of Minnesota, MMC 303, 420 Delaware St. S.E., Minneapolis, Minnesota 55455,
USA \printead{e1}.}
\address[B]{Amy H. Herring is Associate Chair and Professor, Department of
Biostatistics, UNC Gillings School of Public Health, University of North Carolina at Chapel Hill, 3104-D
McGavran-Greenberg Hall, 135 Dauer Drive, Campus Box 7420, Chapel Hill, North Carolina 27599, USA \printead{e2}.}
\end{aug}

%
\begin{abstract}
Alan E. Gelfand was born April 17, 1945, in the Bronx, New York. He
attended public grade schools and did his undergraduate
work at what was then called City College of New York (CCNY, now CUNY),
excelling at mathematics. He then surprised and
saddened his mother by going all the way across the country to Stanford
to graduate school, where he completed his dissertation
in 1969 under the direction of Professor Herbert Solomon, making him an
academic grandson of Herman Rubin and Harold Hotelling. Alan
then accepted a faculty position at the University of Connecticut
(UConn) where he was promoted to tenured associate professor
in 1975 and to full professor in 1980. A few years later he became
interested in decision theory, then empirical Bayes, which
eventually led to the publication of
Gelfand and Smith
[\textit{J. Amer. Statist. Assoc.} \textbf{85} (1990) 398--409], the
paper that introduced the Gibbs sampler to most statisticians
and revolutionized Bayesian computing. In the mid-1990s, Alan's
interests turned strongly to spatial statistics, leading to
fundamental contributions in spatially-varying coefficient models,
coregionalization, and spatial boundary analysis (wombling).
He spent 33 years on the faculty at UConn, retiring in 2002 to
become the James B. Duke Professor of Statistics and Decision Sciences
at Duke University, serving as chair from 2007--2012.
At Duke, he has continued his work in spatial methodology while
increasing his impact in the environmental sciences. To date, he
has published over 260 papers and 6 books; he has also supervised 36
Ph.D. dissertations and 10 postdocs.
This interview was done just prior to a conference of his family,
academic descendants, and colleagues to celebrate his 70th
birthday and his contributions to statistics which took place on April
19--22, 2015 at Duke University.
\end{abstract}

%
\begin{keyword}
\kwd{Bayes}
\kwd{CCNY}
\kwd{Duke}
\kwd{Gibbs sampling}
\kwd{music}
\kwd{spatial statistics}
\kwd{Stanford}
\kwd{UConn}
\end{keyword}
\end{frontmatter}

\section{Early Years, City College, and Stanford}
\textbf{Amy:} Thank you very much for your time and letting us talk with you today.

\textbf{Alan:} I~am delighted!

\textbf{Brad:} You were born in April 1945 just as World War II was ending, went
to the same Bronx, NY junior high school as
George Casella, and bowled and played bridge at CCNY in the 1960s. Tell
us about your parents, your
childhood, your life as a CCNY undergrad, and your path to Stanford for
graduate school.

\begin{figure}

\includegraphics{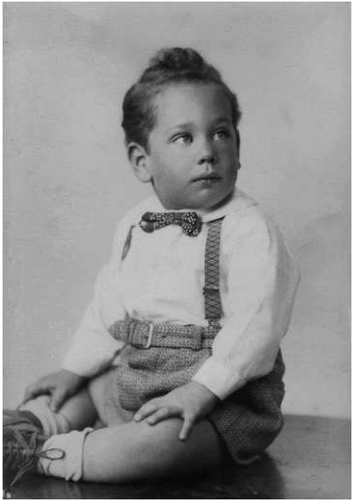}

\caption{Alan, age 2, Fall 1947.}
\label{fig:shorts}
\end{figure}

\textbf{Alan:} I~was ``too young'' all the way through school. At that time
administrators encouraged children to skip grades, and I~graduated high
school and was a freshman in college at 16. Because I~was two years
younger than all the females when I~went off to college,
I~never had much of a social life until I~went out west. I~was really looking for a new experience. In my mind,
California was the land of milk and honey, and it was as far away from
the Bronx as I~could get! I~remember driving away, and my mother
was in tears because she thought I~was going to disappear into the
Pacific and never been seen again!

\textbf{Brad:} What did your father do?

\textbf{Alan:} He was a CPA (Certified Public Accountant), and his fondest
desire was to open Gelfand and Gelfand, CPAs. It was never going to
happen. I~played with numbers too, but not the way he did.

\textbf{Brad:} I~understand that where you grew up in the Bronx was a nice
Jewish family neighborhood.

\textbf{Alan:} Yes, I~grew up in a completely Jewish neighborhood: my elementary
school was 95\% Jewish, the Bronx High
School of Science was 90\% Jewish, and City College was 90\% Jewish. I~thought
the whole world was Jewish! There were many smart kids in NYC, and they
stayed in NYC, went to the specialized high school, and then attended
City College. It
was just the way it was back then, and I~never actually considered
applying anywhere else.

\textbf{Amy:} As a math undergraduate major, what made you choose graduate
school in statistics instead of math?

\textbf{Alan:} This book [the Hogg and Craig text he used at Stanford] is what
opened the door for me; I~just fell in love with mathematical
statistics. I~thought it was so elegant, so cool, all the distribution theory, all
the basic probability theory, the formal inference ideas,
everything about it. I~took mathematical statistics in the beginning of
my senior year and immediately decided it was for me.

\textbf{Brad:} Was your mother heartbroken about your move west?

\textbf{Alan:} She thought it was the end of the world, especially since I~had
full scholarships at Yale and Columbia. It was my decision to go west,
even though my mother tried to bribe me with a car
to stay on the east coast! In the end, I~moved west with two other City
College guys; we roomed together, so I~wasn't totally by myself.

\textbf{Brad:} I~know you are passionate about cars. What did you drive to California?

\textbf{Alan:} I~drove an American Motors Rambler. This car was so slow, it
would do zero to 60 miles per hour in two \emph{minutes}. It was \emph{painful}.
We limped into Palo Alto, and I~remember crossing the Bay Bridge for
the very first time in my life, and suddenly thinking, ``Wow,
San Francisco.'' I~really didn't know how strong a school Stanford was,
or anything about any of the faculty.

However, arriving in Palo Alto in 1965 was just one of those
serendipitous events. It was an incredible time in the sense that a lot
of things were coming together then: the Vietnam War, the
protests, the revolution in music, psychedelia, and drugs. We thought
we were going to change the world. It didn't happen,
but back then there was a spirit that we may never capture again. There
was some innocence in the country that probably is lost
forever. I~particularly embraced the music. You cannot imagine how many
acts I~saw. I~saw the very first public
performances by both Steve Miller and Santana, I~saw Janis Joplin
several times, and I~saw Jefferson Airplane probably a dozen
times. It was wonderful.

\begin{figure}

\includegraphics{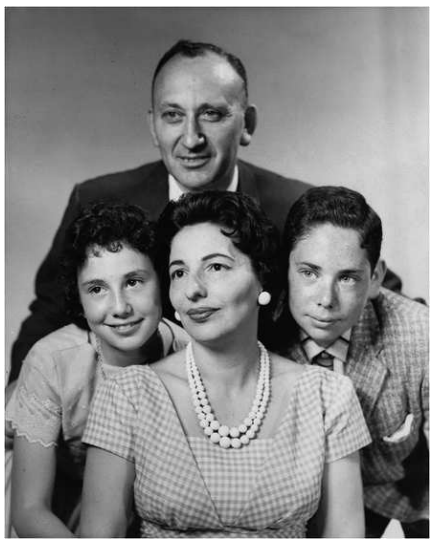}

\caption{Alan (right) with father Abe, mother Frances, and sister Elissa,
just after Alan's high school graduation at age 16, Spring 1961.}
\label{fig:shorts}
\end{figure}

\textbf{Brad:} You're making me crazy; I~play that stuff with my band!

\textbf{Alan:} The face of music just completely changed at that point. Before
then it was Top 40 rock and 3-minute songs, and then all
of a sudden everything opened up; some people claim it was the golden
age for rock and roll. All I~know is it was pretty
exciting.

\textbf{Amy:} When did you first do statistics on a computer?

\textbf{Alan:} Me? I'm still waiting for it to happen! This is an embarrassing
story. My Ph.D. thesis was on seriation
methods: chronological sequencing, particularly driven by
archaeological data. I~proved several theorems about sequencing
data from matrix representations. Then I~had to do a real example,
and\ldots I~hired somebody!

\textbf{Brad:} Tell us about the statistics department at Stanford in the 1960s.

\textbf{Alan:} The faculty was quite prestigious. I~hold the record for the most
courses anybody has ever taken from Charles Stein: 11 quarters. I~also
took the very first
course that Brad Efron taught. He finished his Ph.D. in spring of 1965
and taught that fall. I~had the first year of mathematical
statistics from him. He was inspirational, and I~still have the notes
from that year with him.

I~recall Kai Lai Chung, who would pound chalk to a frazzle; he would go
through a box of chalk in a lecture, in a room filled with chalk dust
and cigarette smoke.
His favorite expression was, ``And we continue to beat the dead horse.''

Of course,
Herb Solomon was my mentor at Stanford, and he was wonderful. He was a
pioneer in
terms of bringing external funding into the department. He had
connections with all the DOD (US Department of Defense) agencies
and with NSF (US National Science Foundation). He raised so much
money that he was providing summer support for a good portion of the
Stanford faculty. He was not adequately
appreciated because they did not view him as a theoretical giant.
However, he was bringing in money at a time when most
statisticians were too pure to get ``dirty'' trying to chase money.

After I~graduated I~went back to Stanford for two decades of summers,
participating in projects with Herb. He was like a second father in
many ways; he and [his wife] Lottie were really very good to me.
I~was young, and he encouraged me to go to Hillel (a worldwide Jewish
campus organization). I~was never religious, but I~went to Hillel
because of the possibility of meeting females.

\textbf{Brad:} Did it work?

\textbf{Alan:} A little bit.

\textbf{Amy:} How did you become interested in statistical applications in
archaeology and law?

\textbf{Alan:} An archaeologist at Stanford raised some quantitative questions
with Herb, and the data were interesting and led to my thesis. Herb had
a real passion for law and justice problems, and in the end this area
was much, much more interesting to me. At first we focused on jury
decision-making, but then we explored various
types of discrimination, jury selection problems, and, eventually,
criminal justice. Later I~also did a fair bit of expert testimony,
which is a very different game from teaching and research.

\textbf{Brad:} You sound like an applied statistician, yet you were not doing
any computing!

\textbf{Alan:} Life wasn't predicated on computing. It was a lot of work just to
invert a $3 \times3$ matrix, so you just didn't do those things. I~did
a lot of analysis
with electric calculators. I~used to have a Monroe and a Frieden on my
desk; these were a step better than those machines where
you turned a crank, a bunch of wheels would roll, and you waited for an
answer to come up. I~didn't really do very much programming or working
with big
computing machines.

\section{UConn, Bayes, the Gibbs Sampler, and Big Data}

\textbf{Brad:} What led you to the University of Connecticut (UConn)?

\textbf{Alan:} I~interviewed at five places: the Stanford Research Institute,
the University of California-Davis, the University of Maryland, Bell Labs,
and UConn. I~decided I~preferred academia. Although UConn was
somewhat sleepy back then, it was close to my family in New York, and
something about New England was appealing, so it emerged as the winner.

\textbf{Amy:} Based on your CV, you went up for tenure at UConn
with just 6 papers: two first-authored papers in the archaeological
literature, a sole-authored paper in \emph{Communications},
two \emph{JASA} papers with your advisor, and a paper in \emph{The
American Statistician}.
How confident you were feeling about this promotion?

\textbf{Alan:} Wow, I~really appreciate that question! I~think there might have
been a few more papers
before tenure. In any event, candidly, I~didn't even know what a good
vita was; all I~knew was that I~was being productive, and it was good
enough, but by today's standards it wouldn't even come close to
``cutting the mustard.'' It was a different time, the bar was different,
and the expectations just weren't what they are today.

I~really had somewhat of a wasted youth. I~was trained to be a
mathematical statistician, but I~was never meant to be a
mathematical statistician.
I~tried to prove theorems because that's what you do if you're a
mathematical statistician, but I~really spent a lot of time
trying to find my niche. I~wandered into decision theory for a while,
which led to a transition to empirical Bayes
(EB).
What eventually emerged was that I~was born to be a stochastic modeler;
it's just that stochastic modeling and, in particular,
hierarchical modeling, didn't really blossom until around 1990. I~was
fortunate to find the area in which I~could contribute, but for the
first 20 years of my career, I~was searching. However, for the last 25
years it has been a wonderful ride, and I~feel very
fortunate.


\textbf{Brad:} You were not ``raised'' as a Bayesian, but you became one of the
world's best-known and strongest advocates for the
Bayesian approach. So I'm intrigued by your ``conversion.'' It sounds
like it was not a dramatic ``Damascus
experience'' like your fellow Stanford grad Jay Kadane, who apparently
had such an
``Oh, what a fool I've been''
moment after a few conversations with Jimmie Savage.
My sense is that your conversion was much more like an empirical
Bayes-style conversion, in
which you put your toe in the water by writing down a mixing
distribution, and pretty soon you find yourself wishing you could
compute posteriors and so forth. Can you tell us about your transition
to Bayesian inference?

\textbf{Alan:} I~was always a likelihoodist,
and I~explored empirical Bayes because of its connections with decision
theory. At the time I~imagined that it would be a nice compromise. But,
of course, it turned out that EB made nobody happy: the
frequentists didn't like it, and the Bayesians didn't either. In EB we
spent a lot of time trying to figure out how to do what Bayesians
eventually could do without needing the corrections that empirical
Bayesians had to develop in order to capture uncertainty.

My full conversion happened in Nottingham. I~took Adrian Smith's short
course at Bowling Green State University in Ohio, which was organized
by Jim Albert. Adrian gave a wonderful week of lectures, and at the end of
that week I~asked, ``Any chance I~could come and spend a sabbatical in
Nottingham?'' And he replied, ``Oh, sure, come!''
He had a numerical integration package called Bayes 4 (\cite{Smietal85}), which could do 6- or 7-dimensional
numerical integrations. That was as cutting edge as you could possibly
imagine back then: sophisticated quadrature
ideas, pseudo-random integration, and a lot of tricks to address the
integration problem in Bayesian inference. I~went
there to see if I~could use his software to solve some empirical Bayes problems.

It's a wonderful story. Adrian picked my family up, all four of us, at
Gatwick Airport.
Adrian rented a rickety old van because he never owned a car (still doesn't).
The very first day in Nottingham, in the space of 24 hours we moved,
bought a car, and went to a barbecue.
Two days later I~went to Nottingham for the first time, and Adrian
suggested I~read \citet{TanWon87}. We decided to explore
variations of their method. A~few weeks later, David Clayton, who was
at Leicester at the time, came to Nottingham for a day, and, in the
context of the Tanner and Wong paper, he remarked that we should read
the paper by
Geman and Geman (\citeyear{GemGem84}) in PAMI (Pattern Analysis and Machine Intelligence, an
IEEE journal). I~remember getting a copy of that paper and
thinking it was clearly much better suited for Bayesian inference than
it was for image reconstruction, which was their
context. The doors had opened, and we saw how to go forward.

You must recall that we were very naive back then. In those days, only
if you were desperate, as a last resort, would
you use Monte Carlo methods. Now such methods are often the \emph{first}
tool, and people don't try to be analytic very often. Whether
that's good or bad, the landscape has certainly changed.

\textbf{Brad:} A great story. Though I~thought Adrian tossed the Geman and Geman
paper in your lap, but in fact he pointed you to Tanner and Wong.

\textbf{Alan:} It was definitely David Clayton who connected us to Geman and Geman,
and David was underappreciated in this regard. He had seen that paper,
and the IEEE journals were a literature that few statisticians read
back then. Also remarkable at the time was Michael Escobar's Ph.D.
thesis, which included what was a Gibbs sampler for implementing
Dirichlet process mixing. He had never heard of the Gibbs sampler; he
just invented this idea for
his particular application. He was also underappreciated.

\textbf{Amy:} One thing that's remarkable about your trajectory is how your
productivity and your creativity have really
increased with age.

\textbf{Alan:} If you look at my vita, I~have about 260 papers now, and maybe
200 of them are post-1990. Two things happened. One is I~found
something I~was reasonably good at,
that created a challenge, and it led me to build interdisciplinary
connections. It just opened up
opportunities that were not there before. Second, as you become more
senior, you are able to build a hierarchy in
your research team, with postdocs, graduate students, and more junior
collaborators. You become
more productive because you have more people helping you to get things
done. It's a different situation from being a junior
researcher where you're much more focused;
these days I'm guiding 10 to 15 different projects.

Finding the Gibbs sampler with Adrian and having that successful paper
was really good fortune. Many
smart people work really hard and don't get so lucky. I~was fortunate
to connect
with a seminal paper, and the only thing I~can congratulate myself for
is the fact that I've worked pretty hard for the
subsequent 25 years in taking advantage of this window of opportunity.
I've been able to keep it growing with students and postdocs and
building bridges. It was such a fantastic opportunity, it was
such a good fit with whatever skill set I~have, so that really is the
best explanation for the delta in productivity.
%
%
%
Again, my eyes really opened up a lot from 1990 forward, and, Brad, you
were on the cusp of it. I~was on sabbatical
while you were finishing your thesis, and I~came back with the Gibbs
sampler, and you lost interest in the thesis! You wanted
to get on board with the Gibbs sampler as much as you could.

\textbf{Brad:} Do you agree with Dennis Lindley's view that Bayes is going to
take over the statistical world,
or do you think the world is going to continue to be kind of a
Bayes-frequentist hybrid, with the choice made out of convenience on
a problem-by-problem basis?

\textbf{Alan:} I~think we all know Dennis forecasted a 21st Bayesian century
because he thought that people would just eventually realize that the
Bayesian paradigm was most natural for inference in
science under uncertainty. But in fact it emerged because it was able
to handle problems that were previously
inaccessible. Moreover, in my mind, it's not in equilibrium yet; we're
still watching an increase in the use of
Bayesian methods. It may be very much according to the type of problem
that you're focusing on; sometimes people say, ``Yes, we need to use
hierarchical modeling and MCMC for this problem, but for
that one, no, maybe we don't.''
I~think usage hasn't actually stabilized yet, and now it's becoming
more complicated with all the big data and data science
that's entering the picture. How will that influence the future of
Bayesian work?
Altogether, it really is becoming a 21st Bayesian century, but
primarily for reasons different from what Lindley might have
liked or envisioned.

\textbf{Brad:} Statisticians are still largely frequentist in what they're
doing. If you submit results of a Phase III clinical trial
to FDA (the US Food and Drug Administration), you still need a
significant $p$-value; many things haven't changed. You're right that
there's a lot of Bayes out there;
for instance, when you go to \surl{amazon.com} to buy an Arnold Schwarzenegger
movie, you also see a link to a
Jean-Claude Van Damme movie. That's the result of a Bayesian inference
engine; it has inferred that you like aging Euro-American action
heroes.
%
%

\begin{figure}

\includegraphics{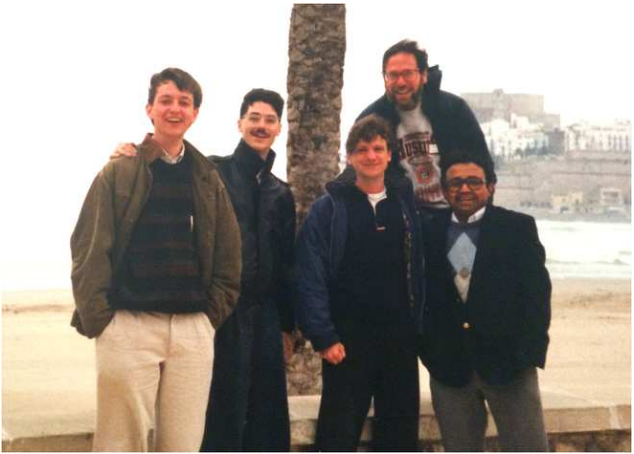}

\caption{L--R: Nick Polson, Brad Carlin, John Wakefield, Alan, and
Dipak Dey on the frigid beach at Pe\~{n}iscola, Spain during the
Valencia 4 meeting, April 1991.}
\label{fig:Val4}
\end{figure}

\textbf{Alan:}
Interestingly, scientists in other fields have no problem thinking in
terms of a Bayesian
paradigm. They're perfectly comfortable inferring what you don't know
given what you've seen, instead of trying to infer
what you might see given what you don't know, which seems backwards.
A lot of the challenge is actually more within the statistical
community itself, and, to date, only certain types
of problems seem to \emph{demand} Bayesian inference.

\textbf{Brad:} MCMC has certainly made the world ``safer'' for being Bayesian.
But are you surprised that nothing has really replaced it? There was a
time when there was a different Bayesian
computational paradigm every 10 years or so, but we've been pretty
stable now for 25
years. Is a new generation of methods going to
replace the current generation of MCMC tools?

\textbf{Alan:} Many say that the size of data sets is going to make MCMC
unusable. I~do think something is going to happen. The candidates
haven't entirely emerged:
INLA (based on integrated nested Laplace approximations) is not
completely satisfying,
ABC (approximate Bayesian computation) certainly has limitations, and
variational Bayes doesn't allow enough inference and is really residing
primarily in the machine learning community. I~don't see sequential
algorithms, particle learning, and particle filters emerging to
overtake MCMC. Still, as data sets keep getting bigger and
bigger, the days when MCMC can still be utilized are going to become
fewer and fewer, so\ldots

\textbf{Brad:} But as computers get faster\ldots

\textbf{Alan:} But the data sets are getting bigger. There's no win in that situation.

\textbf{Brad:} Dueling asymptotics!

\textbf{Alan:} Another concern is what \emph{big data} is about. I~think it's
actually a different philosophy in many situations from
what statistics is about. Most of the work in my world is
hypothesis-driven: I'm thinking about a problem, about a process,
learning about the behavior of the process, and I'm trying to build
models to understand the process, and to hypothesize about
its behavior. But a lot of ``big data analysis'' seems to be searching
big data sets for
structure; you're not hypothesizing much of anything. If statisticians
continue to be interested in hypothesis development and
examination, I'm not sure big data methods are always going to be the answer.

\textbf{Brad:} I~agree; hypothesis investigation requires you to have to have
some idea about uncertainty. You have to have some sort of
variance estimate to test a hypothesis or form a confidence interval,
whereas the big data guys seem primarily interested in a
point estimate or maybe a ranking.

\textbf{Alan:} Statistics must maintain its intellectual genesis, which is
inference under
uncertainty, and continue to argue that such inference is valuable. We
can't live in a purely deductive world, we need a formal
inferential world with randomness.
We have to continue to train people to think that way about problems.

\section{Spatial, Applications, and the Move to Duke}


\textbf{Amy:} In the late 1990s your interests turned strongly to spatial
statistics. How did you become interested in spatial statistics, and
how has it retained your attention for so long?

\textbf{Alan:} A fellow named Mark Ecker came to
UConn for his Ph.D. after earning a master's degree from the University
of Rhode Island. He came into my office one day with that classic spatial
data set on scallop catches in the Atlantic Ocean, and asked, ``What
can I~do with this stuff, and what the heck is a variogram?'' I~said, ``I~have no clue.'' I~had never seen any spatial data, but I~thought it was
interesting. Mark's question literally opened the
door in the spring of 1994, and 20 years later I'm still interested in
spatial statistics. It was just another of those unexpected but
fortunate things that
happened.

At that time, GIS software already permitted visual overlay of spatial
data layers for
making lovely pictures and telling nice descriptive stories, but I~wanted to be able to add an inferential engine to it.
So essentially, Brad, Sudipto Banerjee, and I~set about
creating a fully Bayesian inference engine for spatial analysis; it's
in the book and its revision (\cite{BanCarGel14}). Structured dependence really excited me; I~found it
elegant that you could use it to
learn about the behavior of an
uncountable number of random variables seeing only a finite number of
them. I~enjoyed the
challenges of looking at dependence in two dimensions versus dependence
in one dimension (where there's order), and I~realized
that I~was much more comfortable with interpolation than I~was with
forecasting. I~also realized that there were failures with
the customary asymptotics used with time series, where you let $t$ go
to infinity; that is not what you want to do spatially.
I~got particularly excited about the enormous range of application that
was available as people starting collecting more and
more spatially referenced data. It seemed natural and important to take
advantage of spatial referencing in building models.
I~have been excited to see spatial analysis moving from the periphery
of statistics into the mainstream.

\textbf{Brad:} Sometimes in academia, in order to get a significant raise you
have to threaten to leave for another position. Did you ever think about
leaving the University of Connecticut? You were there for essentially
your whole career; you have had a second career at Duke, but you had a
full career at UConn.

\textbf{Alan:} Definitely, with 33 years at UConn, you are absolutely right.
UConn was always very good to me, and I~felt loyalty and affection for
UConn. They treated me well, and I~thought the quality of life in New
England was good,
so, honestly, I~never really looked.

\textbf{Brad:} There must have been attempts to lure you away?

\textbf{Alan:} Opportunities started becoming serious after 1990; all of a
sudden I~had invitations to become a full professor at a number of
different places---three or four universities in the UK, and maybe
half a dozen in the US.
However, my kids were still finishing high school, and I~wasn't ready
to move.
Duke had contacted me in the mid 1990s and again in the late 1990s;
finally, by 2001 I~was ready, and in 2002 I~made the move.

\textbf{Brad:} \citet{GelSmi90} is clearly your most famous paper, but
what other papers on your CV do you particularly like or feel may have
been underappreciated?

\textbf{Alan:} That's a really good question. I've been pretty lucky, and a lot
of papers have been well-cited
[\emph{note: Alan's h-index at the time of writing is 60}].
Before the spatial work, I~like an underappreciated prior predictive
modeling checks paper
with Dipak Dey, Pantelis Vlachos and Tim Swartz (\cite{Deyetal98}).
Although most of the community
has abdicated this to \emph{posterior} predictive checks
(e.g., \cite{GelMenSte96}), I~think prior predictive checks
have advantages.
Posterior predictive checks are not based on the model that is presumed
to generate the data, and they use the data
twice, making it really hard to criticize models. Prior predictive
checks avoid that trap, and I~don't
understand why there isn't more interest.
I'm revisiting this currently in the context of point patterns to show
how we can better assess
pattern model adequacy.

I~also like our hierarchical centering work for improving MCMC
convergence (Gelfand, Sahu and Carlin, \citeyear{GelSahCar95,GelSahCar96}).
We found a nice analytical solution, at least in Gaussian cases, it was a
demonstrably sensible thing to do, and others continued along those
lines, including Papaspiliopoulos, Roberts and Sk{\"o}ld (\citeyear{PapRobSko07}).

In a different vein, I~think coregionalization is really a lovely idea.
I~couldn't understand why nobody had adopted it as a
general strategy for building multivariate spatial models. I~thought,
what could be easier or more natural
than taking linear transformations of independent processes to create
dependent processes?
The distribution theory works out very well, and the implementations
are also easy (Gelfand et~al., \citeyear{Geletal04}).
This idea is now at the foundation of a lot of \texttt{spBayes} code.

The spatially-varying coefficients paper (\cite{Geletal03})
discusses the remarkable idea that, within the Bayesian
framework, you can learn about spatially-varying intercepts and
spatially-varying slopes as processes without ever actually
observing these processes.
Other papers I~really like include the spatial gradients work I~did
with Sudipto (e.g., Banerjee, Gelfand and Sirmans, \citeyear{BanGelSir03}) and
the wombling papers that subsequently emerged.

\textbf{Amy:} What are your favorite papers focused on applications?

\textbf{Alan:} I'm particularly proud of the species distribution modeling work
that I~did with John Silander and his group at UConn. We presented it
at Carnegie Mellon University at
a Bayesian Case Studies meeting. A~version of it is
in the very first issue of \emph{Bayesian Analysis} (\cite{Geletal06}), and a more technical version (\cite{Geletal05})
was the most cited \emph{JRSS-C} paper of the first decade of the 2000s.
It seems a lot of people from ecology and biological
sciences found it interesting.
At that time, I~was going to South Africa regularly to collaborate.
Researchers were using
simple logistic regressions for presence/absence, which was the state
of the art in the field then. We used a hierarchical model to
induce process features that involve transformation of landscape,
suitability of environments, and availability of
environments. This allowed us to explain not only what you did see but
what you \emph{might} see, with implications for
conservation and management. It resonated well, and I~am still working
on these problems.

Recently I~have gotten into demography, which led to some nice material
with integral projection models (IPMs), particularly arguing
to employ them on the right population scale and again in a fully
hierarchical way.

\textbf{Brad:} Is this how you began collaborating with Jim Clark?

\textbf{Alan:} Yes, and that's another interesting story.
When I~came to interview at Duke, I~went to talk to Jim about
collaborations in Duke's Nicholas School for the Environment.
I~had a simply wonderful two hours with him. He is a real statistician
with a completely appropriate secondary appointment in our department
here at Duke. He imagines
and fits more sophisticated hierarchical models than most statisticians
ever will.

\textbf{Brad:} So you met him the day you interviewed there!

\textbf{Alan:} Yes, I~think we've now reached 40 papers and a book together, so
it's been a wonderful, wonderful
time, and our partnership continues to flourish.

\textbf{Amy:} You have raised an absolutely incredible generation of research
statisticians.
Do you have a strategy for identifying the brightest or most promising students?
What is your mentoring philosophy?

\textbf{Alan:}
I~have never actually recruited students; I~have always just waited for
students to come to me to express interest in working with me. I've
gotten a lot of good students, and my list of ``children'' is really
pretty strong I~think. My primary motivation has been training students
for an academic career. I~think 2/3 to 3/4 of my students are in
academia in some fashion. Not everybody
trains in that fashion, but probably it just reflects the fact that an
academic lifestyle is the best lifestyle I~can
imagine.

As far as developing students, an important aspect is appreciation of
the many ways a modern statistician can
contribute. You can do theory, methodology, modeling, computation, data
analysis, and visualization. You can contribute on many
dimensions, and in fact we try to train across them all. The critical
thing I~try to emphasize to students is to find
what you can really do well and what's going to reward you best. One
size doesn't fit all, and we can't have the same
expectations for every student.

I~also think it's important to encourage fire, passion, and enthusiasm.
We don't do this simply as a 9 to 5 lifestyle, we do
this because we get a lot of satisfaction out of our work. If you're
going to commit a 40-year career to something like this, you've got to
really be in love with it; you don't just do this to pay
the bills.

I~try to foster a fair bit of independence in students because I~think
it's critical that they learn to generate problems and
build their own research agenda. I~do this especially with postdocs,
because they have a two year window and, when they enter
the job market, they need to have a firm sense of what they are going
to do after they get the job.

Also, my style has always been about availability. A lot of faculty are
very structured in the way they interact with
their students, but I've been very flexible. I~sometimes meet with
students at 8~pm just because that's a
good time for me, there's nothing else obligating me, and students
often have ``working in the evening'' lifestyles.
If a student is struggling to do something, I~like to talk about it now
instead of having the student wait for a weekly time slot.

\textbf{Brad:} I~remember when I~was at UConn, you once said, ``Brad, you have
to decide what league you want to play in.''
The implication was,
your work doesn't have to look exactly like mine, or stress mathematics
or computing or any particular tool.
You just have to be in a work environment where you're going to be
productive and where
you're going to be a solid ``player'' in that ``league.''

\textbf{Alan:} That's true, and there are more leagues available now, and more
ways to contribute.
I~think that is what's wonderful about our field.






\begin{figure*}

\includegraphics{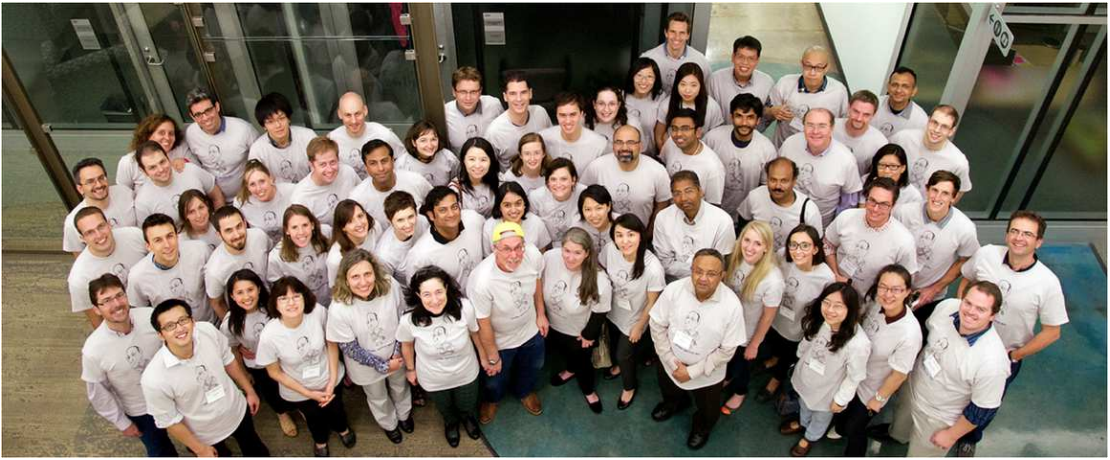}

\caption{Alan (yellow hat) and some of his ``descendants'' and other
guests at the ``G70'' (Gelfand 70th Birthday Conference) poster session,
Duke University, Durham, NC, April 20, 2015.}\label{fig:G70tee}
\end{figure*}

\section{Travel Stories, Hoops, Music, and Future Plans}

\textbf{Amy:} You're also famous for your academic travels. Are there one or two
particularly memorable travel stories you'd like to share?

\textbf{Alan:} Obviously the Valencia meetings have always been a highlight, and
I~was fortunate to make 7 of the 9 Valencia meetings,
and there are too many stories from those to tell. But I~would like to
reminisce about one of the earliest professional meetings
I~went to in Europe. It was when I~was just two years out of my thesis,
1971, and it was an archaeology meeting organized by David Kendall in
Mamaia, Romania on the Black Sea. It was a meeting of statisticians,
applied mathematicians, and archaeologists.
I'd been to Europe before, but I'd never been to a communist country.
There were several remarkable things that happened during this
meeting that will cause it to live in my mind forever.

After arriving in Romania, I~lost my return plane ticket. At that time
there was no
internet. I~sent a telegram to my travel agent back in Storrs, CT to
see if they could help me get a new
ticket. Two days later I~received the following telegram: ``No
information about passenger Blefarx.'' B-L-E-F-A-R-X is how ``Gelfand''
was converted! So I~received no help at all from the travel agency. At
the meeting, they took up a collection for me to pay for my ticket.
I~arrived at the
airport in Bucharest to return to the States, and at the ticket counter
the agent said, ``We have your ticket.''
It apparently had been found on the floor of the terminal in Bucharest
airport and placed at the check-in counter, waiting for
me. Ironically, C.~R. Rao had also lost his ticket, and he and I~were
both part of the collection taken up at the meeting.
It was the first time I~had ever met Professor Rao.

At this same meeting there was a famous Stanford ecologist-statistician
named Luigi Luca Cavalli-Sforza. Luigi Luca came to the meeting
with his wife, who was of noble Italian birth.
The outing for the conference was to take a two-hour bus ride up to the
Danube delta,
where we would get on a boat to travel down the river.
The bus was leaving at 8 o'clock in the morning. I~arrived at roughly
7:55 and said to
the bus driver, ``I~didn't have time to eat anything; can I~run in and
grab something?''
He said, ``No problem, no problem.''
But when I~came back out, the bus was gone. It turned out that also
missing the bus were Luigi Luca and his wife,
who were complaining bitterly.
Thirty minutes later a Mercedes 600 stretch limousine shows up with
Romanian flags on all 4 fenders.
Luigi Luca and his wife climb in\ldots and I~climb in with them; we are
going to catch up!
We went barreling along on these small roads at 120 km/hour,
and after a bit we went right past the bus we were supposed to be on.
We wound up at this cafe near our destination, about 45 minutes ahead
of the bus.
I~will never forget that outrageous ride in a Mercedes limousine on the
back roads of rural Romania.

\textbf{Brad:} Perhaps no statistician in the U.S. is better equipped to answer
this one: Which college basketball program is better: UConn or Duke?

\textbf{Alan:} When I~came to Duke to interview, one of the first things the
Dean said to me was, ``I'll give you a nice parking space, but
don't ask for men's basketball tickets.'' I~said, ``Fine with me!'' I~have gone to many Duke women's basketball games because
I~really like the women's game. But after 33 years at UConn, I~am
afraid I'm always going to be a UConn basketball fan.

\textbf{Brad:} Your music collection is quite famous in some circles; you once
had something like 8000 vinyl record albums.
I~remember finding an original Verve pressing of ``Freak Out'' by Frank
Zappa and the Mothers, and many other rare or obscure albums in your collection.
Can you tell us more about that and your other passions outside of statistics?

\textbf{Alan:} I~started with music back in 1955--1956, with those old, small 42 rpm records
with the fat holes; they had no fidelity whatsoever.
I~listened to the beginnings of rock and roll---Bill Haley and the
Comets, the early Elvis Presley stuff, etc.
In the early 1970s, I~underwent a life-changing event when I~started
collecting jazz. I~collected jazz for probably 25 years. I~had gotten to 6500 vinyl jazz albums comprising a fairly valuable
collection, roughly 8000 pieces of vinyl altogether. Then,
when I~was coming to North Carolina I~had to make a decision: I~was
collecting CD's by that point, what was I~going to do with all the
vinyl? If I~boxed it up, I'd have to find a place to put it when I~got
to Duke, and I~didn't have a place. I~was afraid if I~put
it in storage up north, it might sit there forever; my kids were never
going to be interested in acquiring 6500 pieces of vinyl jazz.
So, I~sold it to a collector in Greenwich Village, New York City.
He came up to Connecticut, packed the collection into 80 boxes, put it
in the back of
a big panel truck, and drove off down the driveway. Before the sale I~had pulled roughly 500 pieces of vinyl that I~thought might never be
available on CD,
and of course that was completely incorrect: now virtually everything
is available online. I~continued to
collect CDs, so now I've got close to 7000 of those dinosaurs. I~probably should have kept the vinyl because vinyl is coming
back, increasing in value, whereas CDs may never come back.

%



\begin{figure}

\includegraphics{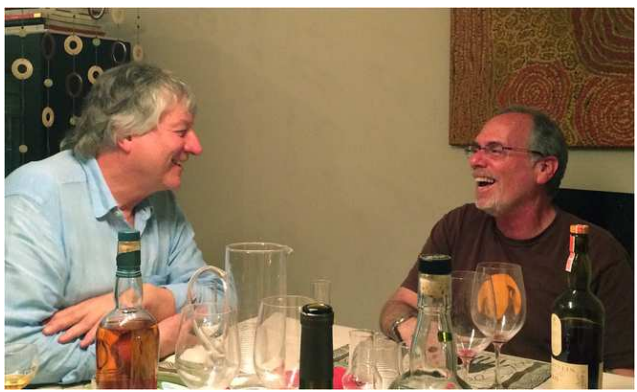}

\caption{Adrian Smith and Alan, pre-G70 dinner, Durham, NC, April 18,
2015.}\label{AlanAdrian}
\end{figure}

Obviously a second passion for me is hoops; I~have always loved
basketball. I~have never had any interest
in American football, and baseball is a bit on the boring side even
though it's quite statistical. I~like soccer a lot, but
basketball is the best game for me.

A third passion for me is cars. I've always flirted with cars, and I~read a couple of car magazines every month.
%
%
A fast, high performance car is probably politically incorrect, but I~still like a quick, good-handling vehicle.
%
%
%

\textbf{Brad:} Now that you are entering your eighth decade on the planet, what
does the future hold for you?
Do you have any major books or other projects under way? Will you
finally open that used vinyl and CD store?

\textbf{Alan:} I'm looking forward to selling my CD collection, but nobody opens
a storefront to do this anymore; I~want to see how well I~can do online using a website called Discogs, or perhaps Amazon.
%
%
%

I~do have three important future commitments. One is
that I~will be an editor for another handbook with Taylor and
Francis/Chapman and Hall, a \emph{Handbook of Environmental
Statistics}. A second thing I'm going to pursue is a project I've
developed called ENMIEP, which is the European Network for Model-Driven
Investigation of
Environmental Processes. I~have a team throughout all of Europe,
including Italy, Portugal, the
UK, Germany, and Spain, and we are trying to find common interests in
environmental research problems. That will be
important because I'm going to be spending a lot of time in Spain (with
my wife, Mariasun Beamonte) and elsewhere in Europe. I~need to have things to do; I~am still curious. However, after turning
70, and after 46 years in the game, maybe it's time to
slow down a bit. My third commitment is to spend as much time as I~can
with the love of my life. She is simply
wonderful, we want to be together, and that has become a priority that
is much more important than publishing a few more papers.
We already have travel plans for Vienna, Prague, Budapest, China, and Africa.
That's really the future.

\textbf{Amy} and \textbf{Brad:} Alan, thank you so much for sharing all of this with us
today! Happy 70th birthday!

\textbf{Alan:} I~have thoroughly enjoyed it. Thank you!





\end{document}